\overfullrule=0pt
\input harvmac

\lref\jusinskas{
   N.~Berkovits and R.~Lipinski Jusinskas,
  ``Light-Cone Analysis of the Pure Spinor Formalism for the Superstring,''
JHEP {\bf 1408}, 102 (2014).
[arXiv:1406.2290 [hep-th]].
}
\lref\AisakaUD{
  Y.~Aisaka, L.~I.~Bevilaqua and B.~C.~Vallilo,
  ``On semiclassical analysis of pure spinor superstring in an $AdS_5$ x $S^5$ background,''
JHEP {\bf 1209}, 068 (2012).
[arXiv:1206.5134 [hep-th]].
}

\lref\brink{
 L.~Brink, M.~B.~Green and J.~H.~Schwarz,
  ``Ten-dimensional Supersymmetric {Yang-Mills} Theory With SO(8) - Covariant Light Cone Superfields,''
Nucl.\ Phys.\ B {\bf 223}, 125 (1983)..
}

\lref\BerkovitsYR{
  N.~Berkovits and O.~Chandia,
  ``Superstring vertex operators in an AdS(5) x S**5 background'',
Nucl.\ Phys.\ B {\bf 596}, 185 (2001).
[hep-th/0009168].
}

\lref\BerkovitsRB{
  N.~Berkovits,
``Covariant quantization of the superparticle using pure spinors,''
JHEP {\bf 0109}, 016 (2001).
[hep-th/0105050].
}

\lref\GalperinAV{
  A.~Galperin, E.~Ivanov, S.~Kalitsyn, V.~Ogievetsky and E.~Sokatchev,
``Unconstrained N=2 Matter, Yang-Mills and Supergravity Theories in Harmonic Superspace,''
Class.\ Quant.\ Grav.\  {\bf 1}, 469 (1984)..
}

\lref\vallilo{
  B.~Vallilo and L.~Mazzucato,
  ``The Konishi Multilpet at Strong Coupling,''
JHEP {\bf 1112}, 029 (2011).
[arXiv:1102.1219 [hep-th]].
}

\lref\MikhailovAF{
  A.~Mikhailov,
 ``Finite dimensional vertex,''
JHEP {\bf 1112}, 005 (2011).
[arXiv:1105.2231 [hep-th]].
}

\lref\mikh{
  A.~Mikhailov and R.~Xu, ``BRST cohomology of the sum of two pure spinors,''
to appear.
}

\lref\minahan{
  J.~Minahan,
  ``Holographic three-point functions for short operators,''
JHEP {\bf 1207}, 187 (2012).
[arXiv:1206.3129 [hep-th]].
}

\lref\BerkovitsGA{
  N.~Berkovits,
  ``Simplifying and Extending the AdS(5) x S**5 Pure Spinor Formalism,''
JHEP {\bf 0909}, 051 (2009).
[arXiv:0812.5074 [hep-th]].}

\lref\BerkovitsMaz{
  N.~Berkovits and L.~Mazzucato,
`` Taming the b antighost with Ramond-Ramond flux,''
JHEP {\bf 0111},  019 (2010).
[arXic:1004.5140 [hep-th].}


\lref\BerkovitsBT{
  N.~Berkovits,
  ``Pure spinor formalism as an N=2 topological string'',
JHEP {\bf 0510}, 089 (2005).
[hep-th/0509120].}

\lref\BerkovitsXU{
  N.~Berkovits,
 ``Quantum consistency of the superstring in AdS(5) x S**5 background,''
JHEP {\bf 0503}, 041 (2005).
[hep-th/0411170].
}


\lref\BerkovitsFE{
  N.~Berkovits,
  ``Super Poincare covariant quantization of the superstring'',
JHEP {\bf 0004}, 018 (2000).
[hep-th/0001035].
}


\lref\MazzucatoJT{
  L.~Mazzucato,
  ``Superstrings in AdS'',
[arXiv:1104.2604 [hep-th]].
}


\lref\HeslopNP{
  P.~Heslop and P.~S.~Howe,
  ``Chiral superfields in IIB supergravity'',
Phys.\ Lett.\ B {\bf 502}, 259 (2001).
[hep-th/0008047].
}







\lref\SohniusWK{
  M.~F.~Sohnius,
  ``Bianchi Identities for Supersymmetric Gauge Theories,''
Nucl.\ Phys.\ B {\bf 136}, 461 (1978).
}

\lref\ArutyunovGA{
  G.~Arutyunov and S.~Frolov,
  ``Foundations of the $AdS_5 \, x \, S^5$ Superstring. Part I,''
J.\ Phys.\ A {\bf 42}, 254003 (2009).
[arXiv:0901.4937 [hep-th]].
}

\lref\MetsaevIT{
  R.~R.~Metsaev and A.~A.~Tseytlin,
  ``Type IIB superstring action in AdS(5) x S**5 background,''
Nucl.\ Phys.\ B {\bf 533}, 109 (1998).
[hep-th/9805028].
}



\lref\HoweSRA{
  P.~S.~Howe and P.~C.~West,
  ``The Complete N=2, D=10 Supergravity'',
Nucl.\ Phys.\ B {\bf 238}, 181 (1984).
}

\lref\BerkovitsULM{
  N.~Berkovits,
  ``Sketching a Proof of the Maldacena Conjecture at Small Radius,''
[arXiv:1903.08264 [hep-th]].
}
\lref\GomezSLA{
  H.~Gomez and C.~R.~Mafra,
  ``The closed-string 3-loop amplitude and S-duality,''
JHEP {\bf 1310}, 217 (2013).
[arXiv:1308.6567 [hep-th]].
}
\lref\FrolovAV{
  S.~Frolov and A.~A.~Tseytlin,
  ``Semiclassical quantization of rotating superstring in AdS(5) x S**5,''
JHEP {\bf 0206}, 007 (2002).
[hep-th/0204226].
}
\lref\ChoNFN{
  M.~Cho, S.~Collier and X.~Yin,
  ``Strings in Ramond-Ramond Backgrounds from the Neveu-Schwarz-Ramond Formalism,''
[arXiv:1811.00032 [hep-th]].
}
\lref\vallilo{
O.~Chandia and B.~Vallilo,
``A superfield realization of the integrated vertex operator in an $AdS_5\times S^5$ background,''
JHEP {\bf 1710}, 178 (2017).
[arXiv:1709.00517 [hep-th]].
}

\lref\MinahanFH{
  J.~A.~Minahan,
  ``Holographic three-point functions for short operators,''
JHEP {\bf 1207}, 187 (2012).
[arXiv:1206.3129 [hep-th]].
}
\lref\MazzucatoJaT{
  L.~Mazzucato,
  ``Superstrings in AdS,''
Phys.\ Rept.\  {\bf 521}, 1 (2012).
[arXiv:1104.2604 [hep-th]].
}

\lref\BerkovitsPS{
  N.~Berkovits and T.~Fleury,
  ``Harmonic Superspace from the $AdS_5\times S^5$ Pure Spinor Formalism,''
JHEP {\bf 1303}, 022 (2013).
[arXiv:1212.3296 [hep-th]].
}
\lref\BerkovitsULM{
  N.~Berkovits,
  ``Sketching a Proof of the Maldacena Conjecture at Small Radius,''
[arXiv:1903.08264 [hep-th]].
}
\lref\BedoyaQZ{
  O.~A.~Bedoya, L.~I.~Bevilaqua, A.~Mikhailov and V.~O.~Rivelles,
  ``Notes on beta-deformations of the pure spinor superstring in AdS(5) x S(5),''
Nucl.\ Phys.\ B {\bf 848}, 155 (2011).
[arXiv:1005.0049 [hep-th]].
}

\lref\NBdynamical{
N.~Berkovits,
``Dynamical twisting and the b ghost in the pure spinor formalism,''
JHEP 06, 091 (2013)
[arXiv:1305.0693 [hep-th]].}

\lref\NBcov{
N.~Berkovits,
``Covariant Map Between Ramond-Neveu-Schwarz and Pure Spinor Formalisms for the Superstring,''
JHEP 04, 024 (2014)
[arXiv:1312.0845 [hep-th]].}

\lref\NBuntwist{
N.~Berkovits,
``Untwisting the pure spinor formalism to the RNS and twistor string in a flat and AdS$_{5} \times$ S$^{5}$ background,''
JHEP 06, 127 (2016)
[arXiv:1604.04617 [hep-th]].}

\lref\NBsusy{
N.~Berkovits,
``Manifest spacetime supersymmetry and the superstring,''
JHEP 10, 162 (2021)
[arXiv:2106.04448 [hep-th]].}

\lref\NBC{
N.~Berkovits, O. Chandia, J. Gomide and L.N.S. Martins,
``B-RNS-GSS heterotic string in curved backgrounds,'' [arXiv:2211.06899 [hep-th]].}

\lref\nekrasov{
N.~Nekrasov, private communication.}

\lref\baulieuone{
L.~Baulieu,
``SU(5)-invariant decomposition of ten-dimensional Yang-Mills supersymmetry,''
Phys. Lett. B 698, 63-67 (2011)
[arXiv:1009.3893 [hep-th]].}

\lref\costello{K.~Costello and S.~Li,
``Quantization of open-closed BCOV theory, I,''
[arXiv:1505.06703 [hep-th]].}

\lref\NBufive{
N.~Berkovits,
``Quantization of the superstring with manifest U(5) super-Poincare invariance,''
Phys. Lett. B 457, 94-100 (1999)
[arXiv:hep-th/9902099 [hep-th]].}

\lref\NBooguri{
N.~Berkovits, H.~Ooguri and C.~Vafa,
``On the world sheet derivation of large N dualities for the superstring,''
Commun. Math. Phys. 252, 259-274 (2004)
[arXiv:hep-th/0310118 [hep-th]].}

\lref\vertical{
A.~Sen,
``Off-shell Amplitudes in Superstring Theory,''
Fortsch. Phys. 63, 149-188 (2015)
[arXiv:1408.0571 [hep-th]].}

\lref\baulieu{
L.~Baulieu,
``Transmutation of pure 2-D supergravity into topological 2-D gravity and other conformal theories,''
Phys. Lett. B 288, 59-68 (1992)
doi:10.1016/0370-2693(92)91954-8
[arXiv:hep-th/9206019 [hep-th]].}

\lref\berknek{N.~Berkovits and N.~Nekrasov,
``Multiloop superstring amplitudes from non-minimal pure spinor formalism,''
JHEP 12, 029 (2006)
[arXiv:hep-th/0609012 [hep-th]].}

\lref\topological{
N.~Berkovits,
``Pure spinor formalism as an N=2 topological string,''
JHEP 10, 089 (2005)
[arXiv:hep-th/0509120 [hep-th]].}

\lref\siegelqm{
W.~Siegel,
``Classical Superstring Mechanics,''
Nucl. Phys. B 263, 93-104 (1986).}

\lref\sorokin{D.~P.~Sorokin,
``Superbranes and superembeddings,''
Phys. Rept. 329, 1-101 (2000)
[arXiv:hep-th/9906142 [hep-th]].}

\def\bar{\overline}

\def\a{{\alpha}}

\def\l{{\lambda}}
\def\lb{{\overline\lambda}}

\def\wb{{\overline w}}
\def\ub{{\overline u}}

\def\lb{{\overline\lambda}}
\def\b{{\beta}}

\def\g{{\gamma}}
\def\G{{\Gamma}}
\def\Gb{{\overline\Gamma}}

\def\d{{\delta}}
\def\e{{\epsilon}}
\def\s{{\sigma}}
\def\k{{\kappa}}

\def\L{{\Lambda}}
\def\Lb{{\overline\Lambda}}

\def\Wb{{\overline W}}

\def\Vb{{\overline V}}
\def\ub{{\overline u}}
\def\vb{{\overline v}}

\def\O{{\Omega}}
\def\Ob{{\overline\Omega}}

\def\half{{1\over 2}}
\def\p{{\partial}}

\def\pb{{\overline\partial}}
\def\t{{\theta}}

\def\etawt{{\widetilde \eta}}
\def\etat{{\widetilde \eta}}
\def\xit{{\widetilde \xi}}
\def\xiwt{{\widetilde \xi}}
\def\gt{{\widetilde \g}}
\def\gwt{{\widetilde \g}}
\def\bt{{\widetilde \b}}
\def\bwt{{\widetilde \b}}
\def\phit{{\widetilde \phi}}
\def\phiwt{{\widetilde \phi}}

\def\T{{\Theta}}

\def\S{{\Sigma}}
\def\Sb{{\overline \S}}

\def\ub{{\overline{u}}}
\def\lb{{\overline{\lambda}}}

\Title{\vbox{\baselineskip12pt
\hbox{}}}
{{\vbox{\centerline{D=5 Holomorphic Chern-Simons and the}
\smallskip
\centerline{ Pure Spinor Superstring}}} }
\bigskip\centerline{Nathan Berkovits\foot{e-mail: nathan.berkovits@unesp.br}}
\bigskip
\centerline{\it ICTP South American Institute for Fundamental Research}
\centerline{\it Instituto de F\'\i sica Te\'orica, UNESP - Univ. 
Estadual Paulista }
\centerline{\it Rua Dr. Bento T. Ferraz 271, 01140-070, S\~ao Paulo, SP, Brasil}
\bigskip

\vskip .1in

The physical states of D=5 holomorphic Chern-Simons theory correspond to on-shell D=10 open superstring states in the cohomology of $q_+$, where $q_+$ is one of the 16 spacetime supersymmetry generators. Scattering amplitudes of these states can be computed either using the usual Ramond-Neveu-Schwarz (RNS) superstring  prescription with N=1 worldsheet supersymmetry, or using a topological  $\hat c$=5 string theory with twisted N=2 worldsheet supersymmetry. 

It will be argued that the relation between D=5 holomophic Chern-Simons and the RNS superstring is identical to the relation between the the pure spinor superstring and the recently constructed B-RNS-GSS superstring which has both N=1 worldsheet supersymmetry and D=10 spacetime supersymmetry. Physical states of the pure spinor superstring correspond to on-shell B-RNS-GSS states which are in the cohomology of $\l^\a q_\a$, where $\l^\a$ is a D=10 pure spinor. And scattering amplitudes of these states can be computed either using the full B-RNS-GSS superstring prescription with N=1 worldsheet supersymmetry,  or using the pure spinor superstring amplitude prescription with twisted N=2 worldsheet supersymmetry. This should be useful for proving equivalence of the RNS and pure spinor amplitude prescriptions and for clarifying the relation of their multiloop subtleties.

\vskip .1in

\Date {November 2022}
\newsec{Introduction}

The RNS formalism for the superstring has a beautiful geometric foundation coming from the gauge-fixing
of an N=(1,1) worldsheet super-reparameterization invariant action, but the lack of manifest spacetime supersymmetry
complicates the computation of multiloop scattering amplitudes and prevents the description of Ramond-Ramond backgrounds.
On the other hand, the pure spinor formalism for the superstring has manifest spacetime supersymmetry and can describe Ramond-Ramond backgrounds, but lacks a geometric
foundation where the pure spinor BRST operator comes from gauge-fixing worldsheet reparameterization invariance. 
Unlike the N=1 worldsheet supersymmetric RNS formalism which requires summing over spin structures, the pure spinor formalism has twisted N=2 worldsheet supersymmetry and there is no need to sum over spin structures. There is complete agreement between all scattering amplitudes which have been computed using the two formalisms, but equivalence has not yet been proven and multiloop amplitude computations in the two formalisms have different types of subtleties.

In previous papers \NBdynamical\NBcov\NBuntwist\NBsusy\NBC, the RNS and pure spinor formalisms have been related to
a worldsheet action with N=(1,1) worldsheet super-reparameterization invariance and manifest D=10 spacetime supersymmetry.
In addition to the usual bosonic worldsheet superfields $X^m$ of the RNS formalism which are spacetime vectors, the action includes a fermionic worldsheet superfield $\Theta^\a$ which is a spacetime spinor. In this paper, it will be useful to also include
a bosonic worldsheet superfield $\Vb_\a$
which is a spacetime spinor of the opposite chirality and plays the role of the non-minimal variables in the pure spinor formalism.
For the Type II superstring, one has both left and right-moving versions of the additional
worldsheet superfields which are related to each other in the usual way for the open superstring. For simplicity, we will only discuss the open superstring in this paper and will ignore the right-moving versions of the additional worldsheet superfields.

In a flat background, the worldsheet action is 
\eqn\action{S =- \int d^2 z \int d^2 \k [\half  D X^m \overline D X_m + \Phi_\a \overline D \T^\a + 
\Wb^\a \overline D\Vb_\a] }
where $D = {\p\over{\p\k}} - \k \p_z$ and $\overline D ={\p\over{\p\overline\k}} + \overline\k \p_{\overline z}$ are the fermionic N=(1,1) derivatives, $(\Phi_\a, \Wb^\a)$ are conjugate momenta superfields to $(\T^\a, \Vb_\a)$, and the D=10 spacetime supersymmetry transformations are 
\eqn\susyone{\d\T^\a = \e^\a, \quad \d X^m = -\half \e\g^m \T, \quad \d \Phi_\a = \half (\e\g^m)_\a DX_m - {1\over 8} (\e\g^m \T)(\g_m D\T)_\a.}
After integrating out auxiliary fields, the action of \action\ in terms of components is
\eqn\actioncomp{S = \int d^2 z [\half \p x^m \pb x_m +\half \psi^m \pb \psi_m +  p_\a \pb \t^\a + \O_\a \pb \L^\a + \Ob^\a\pb \Lb_\a + S^\a \pb R_\a ]}
where, ignoring auxiliary components of the superfields, $X^m = x^m + \k \psi^m$, $\T^\a = \t^\a + \k \L^\a$, $\Phi_\a = \O_\a + \k p_\a$, 
$\Vb_\a = -\Lb_\a + \k R_\a$, $\Wb^\a = S^\a + \k\Ob^\a$.

Note that worldsheet supersymmetric actions for the superstring including the $\T^\a$
twistor-like superfield were proposed in the superembedding approach reviewed in \sorokin, however, quantization in this approach was complicated by the presence of second-class constraints. As first discussed by Siegel \siegelqm,
replacement of the second-class constraints with first-class constraints requires inclusion of the conjugate momentum superfield $\Phi_\a$. 

Since the worldsheet and spacetime supersymmetric action of \action\ shares features of the pure spinor formalism, the RNS formalism and the Green-Schwarz-Siegel formalism, 
it is called the B-RNS-GSS action. Because of opposite statistics, the contribution of the extra worldsheet superfields to the worldsheet conformal anomaly cancels. But because of these extra worldsheet superfields, the usual RNS physical
spectrum is a subset of physical states in the B-RNS-GSS formalism. As will be shown here, this subset of states in the B-RNS-GSS formalism is
analogous to the subset of physical open superstring states in the RNS formalism corresponding to D=5 holomorphic Chern-Simons states. Furthermore, just as superstring amplitudes can be computed using either the N=1 worldsheet supersymmetric RNS formalism or the twisted N=2 worldsheet supersymmetric pure spinor formalism, D=5 holomorphic Chern-Simons amplitudes can be computed using either N=1 or twisted N=2 worldsheet supersymmetric prescriptions.

As discussed in \nekrasov\baulieuone\costello,  holomorphic Chern-Simons in five complex dimensions describes the subset of super-Yang-Mills states in ten Euclidean dimensions which are annihilated by the
spacetime supersymmetry generator $q_+$ where, under the $SU(5)\times U(1)$ decomposition of $SO(10)$, $q_+$ is the $SU(5)$-invariant component of the spacetime
supersymmetry generator $q_\a$ with U(1)-charge $+{5\over 2}$. More precisely, holomorphic Chern-Simons states need to be annihilated by $q_+$ up to a BRST gauge transformation, i.e. the states should be 
in the cohomology of $(Q-q_+)$ where $Q$ is the usual BRST operator of the RNS formalism. 
In the RNS formalism, the spacetime supersymmetry generator in the $-\half$ and $+\half$ picture is
\eqn\susyrns{ q^{-\half}_\a= \oint e^{-{\phi \over 2}} \Sigma_\a, \quad 
q^{+\half}_\a= [Q, \oint \xi e^{-{\phi \over 2}} \Sigma_\a ], } 
where $\Sigma_\a$ and $\Sigma^\a$ are spin fields of $5\over 8$ conformal weight constructed from $\psi^m$, and $\beta = \p\xi e^{-\phi}$ and $\g = \eta e^\phi$ are the RNS superconformal ghosts. Furthermore, physical states in the RNS formalism are required to be in the ``small" Hilbert space independent of the $\xi$ zero mode, i.e. they need to be annihilated by $\eta_0\equiv \oint \eta$. 

So D=5 holomorphic Chern-Simons states can be defined as the subset of RNS states in the cohomology of $(Q-q^{+\half}_+)$ which are annihilated by $\eta_0$. Since 
\eqn\simAA{e^B e^A (Q-q_+^{+\half})e^{-A}e^{-B} = e^{ B} Q e^{-B}= Q, }
$$e^B e^A (\eta_0)e^{-A}e^{-B} = e^B(\eta_0 + q^{-\half}_+) e^{-B} =   q^{-\half}_+$$
where $A=\oint  e^{-{\phi \over 2}} \Sigma_+\xi$ and $B =\oint  e^{{\phi \over 2}} \Sigma^+ \eta$ , 
these states can equivalently be defined as the subset of RNS states in the cohomology of $Q$ which are annilhilated by  $q^{-\half}_+$. 

As will be shown here, $g$-loop scattering amplitudes of these D=5 holomorphic Chern-Simons states can be computed in two manners. The first method is the usual RNS amplitude prescription using RNS vertex operators for these states integrated on an N=1 superconformal Riemann surface of genus $g$ with the appropriate number of RNS picture-raising operators. However, when all states are holomorphic Chern-Simons states, there is a second method involving a ``topologically twisted" amplitude prescription in which all worldsheet variables carry integer conformal weight and the N=1 superconformal generator $G$ of the RNS formalism is split into $G= G^+ + G^-$ where $G^+$ and $G^-$ are twisted N=2 superconformal generators carrying $+1$ and $+2$ conformal weight respectively.

To topologically twist \baulieu\NBcov,  split the 10 $\psi^m$ worldsheet variables into 5 complex pairs $(\psi^a, \psi_a)$ for $a=1$ to 5 and define $\G^a = \g \psi^a $ and $\Gb_a = \g^{-1}\psi_a$ with conformal weight 0 and 1. 
If the $(\b,\g)$ ghosts are replaced with ``tilded" ghosts $(\gwt,\bwt)$ of conformal weight $(-1,+2)$ defined as 
\eqn\tildedghosts{\widetilde \g = (\g)^2 = \eta\p\eta e^{2\phi}, \quad \bwt = (2\g)^{-1}\b + (2\g^2)^{-1} \psi^a \psi_a,} 
this twisting procedure does not affect the conformal anomaly since the $\psi^m\to(\G^a,\Gb_a)$ twisting shifts $c=5$ to $c=-10$, and the $(\g,\b)\to(\gwt,\bwt)$ twisting shifts $c=11$ to $c=26$. 
Fermionizing $\gwt = \etawt e^{\phiwt}$ and $\bwt = \p\xiwt e^{-\phiwt}$ and defining $J=\p H = \psi^a \psi_a$, one finds that
\eqn\fermtilde{\etat= e^{-\half(\phi +H)} = e^{-{\phi \over 2}} \Sigma_+, \quad\xit=e^{\half(\phi +H)} =e^{{\phi \over 2}} \Sigma^+.}
So all holomorphic Chern-Simons states are in the small ``tilded" Hilbert space since they are annihilated by
$q_+^{ -\half} = \oint \etat$.

In terms of these twisted worldsheet variables, the RNS BRST operator is
\eqn\twbrst{Q = \oint [G^+ + \gwt (G^ - + b) + c (T- \bt\p\gt- \p(\bt \gt) - b\p c)]}
where $G^+ = \p x_a \G^a$, $G^- = \p x^a \Gb_a$ and $T = -\p x_a \p x^a - \Gb_a \p\G^a$. Since 
\eqn\simpq{Q = e^{- \oint (c G^- +\bwt c \p c)} \oint (\p x_a \G^a + \gwt b)~  e^{ \oint (c G^- +\bwt c \p c)},}
states in the cohomology of $Q$ are related by a similarity transformation to states in the cohomology of $\oint \p x_a \G^a$ that are independent of $(b,c)$ and $(\gwt,\bwt)$, which is the standard definition of D=5 holomorphic Chern-Simons states. And the topologically twisted amplitude prescription can be used to compute $g$-loop scattering amplitudes by integrating holomorphic Chern-Simons vertex operators on a genus $g$ bosonic Riemann surface, together with $(3g-3)$ insertions of $G^-$ sewed with the Beltrami differentials. 

Since the amplitudes are independent of the picture-changing operator locations, one can show that the RNS prescription agrees with this topologically twisted prescription by choosing the picture-changing operators of the RNS prescription to be located on top of the $b$ ghost insertions. The only subtlety is that the similarity transformation of \simAA\ does not commute with the picture-changing operator since
$e^{B}e^A \xi e^{-A} e^{-B} = \xi +  \xiwt$. So the RNS picture-changing operator $Q(\xi)$ is transformed into 
\eqn\newpco{Q(\xi + \xiwt) = Q(\xi) + e^{\phiwt} (G^- + b).} 
The first term $Q(\xi)$ in \newpco\ can be ignored in the small ``tilded" Hilbert space since $\xi$ is annihilated by $\etawt_0$. And when inserted on the $b$ ghost insertion, the second term in \newpco\ produces $b e^{\phiwt} G^-$ which, after integrating out the $(c,b)$ and $(\gwt, \bwt)$ variables, reduces to the $G^-$ insertion of the twisted topological prescription.  

Just as holomorphic Chern-Simons states are a subset of states in the BRST cohomology of the RNS formalism whose scattering amplitudes can be computed using either the usual N=1 prescription or the topologically twisted prescription, it will be argued in this paper that there is a subset of states in the BRST cohomology of the B-RNS-GSS formalism whose  scattering amplitudes can be computed using either an N=1 prescription or a topologically twisted prescription.
The BRST operator $Q$ in the B-RNS-GSS formalism is constructed in the usual manner from N=1 superconformal generators built out of the superfields $(X^m, \T^\a, \Phi_\a, \Vb_\a, \Wb^\a)$ of \action\ together with the usual N=1 superconformal ghosts. And the BRST cohomology in the small Hilbert space defined by $QV=\eta_0 V=0$ contains the usual RNS physical open superstring states depending only on the $X^m$ superfield, as well as extra states with nontrivial dependence on the superfields $(\T^\a, \Phi_\a, \Vb_\a, \Wb^\a)$.  

However, just as $QV = q_+^{-\half} V=0$ defines a subset of states in the RNS formalism corresponding to D=5 holomorphic Chern-Simons states, the conditions
$QV =  \etawt_0 V=0$ in the B-RNS-GSS formalism will define a subset of states corresponding to the usual open superstring spectrum. 
But instead of defining $\etawt_0$ to be the $SU(5)$-covariant $q^{-\half}_+= \oint e^{-{\phi\over 2}}\S_+$ of \fermtilde, one can use the spacetime spinors $(\L^\a, \Lb_\a, R_\a)$ of the
B-RNS-GSS formalism to define an $SO(10)$-covariant $\etawt_0$ as
\eqn\etapure{\widetilde\eta_0 = \oint e^{-{\phi \over 2}}( :\l^\a \Sigma_\a:  -(\l\g^{mnp} \S) {{\lb\g_{mnp}r}\over{8(\l\lb)^2}}) }
where $(\l^\a, \lb_\a)$ are $D=10$ pure spinors satisfying $\l\g^m\l =\lb\g^m \lb=0$ and $r_\a$ is a fermionic spinor satisfying $\lb\g^m r =0$. The first term in \etapure\ coincides with \fermtilde\ when $\l^\a$ is constant and the second term in \etapure\ is necessary so that $\{Q, \etat_0\}=0$.
The constrained spinors $(\l^\a, \lb_\a, r_\a)$ in \etapure\ are defined in terms of $(\L^\a, \Lb_\a, R_\a)$ through the relations
\eqn\rellambda{\L^\a = \l^\a + u^m {{(\g_m \lb)^\a}\over{2(\l\lb)}}, \quad 
\Lb_\a = \lb_\a + \ub_m {{(\g^m \l)_\a}\over{2(\l\lb)}}, \quad R_\a = r_ \a + \rho_m {{(\g^m \l)_\a}\over{2(\l\lb)}} }
where $(u^m, \bar u_m, \rho_m)$ are spacetime vectors which each have 5 independent components.

To topologically twist the B-RNS-GSS formalism, shift the B-RNS-GSS stress tensor as $T\to T+\half \p J$ where
\eqn\brnsjj{ J = -\L^\a \O_\a + R_\a S^\a - {{(\l\g^m \g^n \lb)}\over {2(\l\lb)}} \psi_m \psi_n - {{(\lb\g^m \g^n\g^p r)}\over {4(\l\lb)^2}} \psi_m \psi_n\psi_p.}
This $U(1)$ generator satisfies $[\oint J, G] = G^+ - G^-$ where $G$ is the $N=1$ superconformal generator of the B-RNS-GSS
formalism and $(J, G^+,  G^-, T)$ form a twisted $N=2$ algebra. Furthermore, if one defines $\gt$ and $\bt$ as in \tildedghosts,
\etapure\ can be expressed as $\etat = e^{-\half (\phi+ H)}$ where $J = \p H.$
In terms of the twisted $N=2$ superconformal generators and tilded ghosts, the B-RNS-GSS BRST operator has the same form as \twbrst\
where
$G^+$ is the non-minimal pure spinor BRST current and $G^-$ and $T$ are the composite $B$ ghost and stress tensor of the non-minimal pure spinor formalism. So using the same arguments as for holomorphic Chern-Simons, B-RNS-GSS states satisfying
$QV = \etat_0 V=0$ are equivalent to superstring states in the cohomology of the pure spinor BRST operator. 

Furthermore, as in the amplitude computation for holomorphic Chern-Simons states, one should be able to use either the $N=1$ superconformal prescription or the topologically twisted $N=2$ prescription to compute $g$-loop amplitudes of these open superstring states. The $N=1$ superconformal prescription corresponds to the usual RNS amplitude prescription in which RNS vertex operators are integrated on an $N=1$ super-Riemann surface with an appropriate number of picture-changing operators, and where the functional integral over the $(\T^\a, \Phi_\a)$ superfields cancels the functional integral over the $(\Vb_\a, \Wb^\a)$ superfields. And the topologically twisted $N=2$ amplitude prescription corresponds to the pure spinor amplitude prescription in which vertex operators in the cohomology of the pure spinor BRST operator are integrated on a genus $g$ bosonic Riemann surface, together with $(3g-3)$ insertions of the composite pure spinor $B$ ghost sewed with the Beltrami differentials. 

As in the holomorphic Chern-Simons amplitudes, showing the equivalence of these prescriptions requires that the $N=1$ superconformal prescription is independent of the location of the picture-changing operators and that the BRST-trivial term $Q(\xi)$ in \newpco\ can be ignored. In the holomorphic Chern-Simons case, one can easily verify that these two requirements are satisfied. However, in general superstring multiloop amplitudes, the first requirement is not satisfied since changing the locations of the picture-changing operators can produce surface terms. Furthermore,  when expressed in terms of the non-minimal pure spinor variables, the RNS variable $\xi$ contains inverse powers of $(\l\lb)$. In the non-minimal pure spinor formalism, BRST-trivial quantities can contribute to scattering amplitudes if they contain enough inverse powers of $(\l\lb)$. So for general multiloop amplitude computations in which the composite $B$ ghost also contributes inverse powers of $(\l\lb)$, the BRST-trivial term $Q(\xi)$ cannot be ignored and the second requirement is not satisfied. 

Because of these two types of subtleties, the topologically twisted prescription in which the composite pure spinor $B$ ghost is sewed with Beltrami differentials can only be used for special superstring multiloop amplitudes in which at least one spacetime supersymmetry is preserved by the external states. As explained in \berknek, pure spinor computations of these ``F-term" scattering amplitudes do not have any subtleties since the amplitude integrands are independent of the picture-changing operator locations and there are not enough inverse powers of $(\l\lb)$ to cause a breakdown of BRST invariance. However, for ``D-term" superstring multiloop amplitudes in which all spacetime supersymmetries are broken by the external states, the topologically twisted prescription needs to be modified in order to preserve BRST invariance. It would be very interesting to determine the explicit form of these modifications coming from the dependence on picture-changing operator
locations and $Q(\xi)$ terms in the B-RNS-GSS formalism.
 
In section 2 of this paper, D=5 holomorphic Chern-Simons states will be described as a subset of open superstring states, and scattering amplitudes will be computed using both
the usual RNS prescription and the twisted topological prescription. 
In section 3, the B-RNS-GSS formalism with both worldsheet and spacetime supersymmetry will be defined and the usual superstring states will be argued to correspond to BRST-invariant states in this formalism which are annilihated by $\etat_0$ of \etapure. Scattering amplitudes of these states can be computed using either an $N=1$ or twisted $N=2$ worldsheet supersymmetric prescription, and some comments are made on how this could be used to prove equivalence of the RNS and pure spinor amplitude prescriptions.

\newsec{D=5 Holomorphic Chern-Simons Theory}

In this section, D=5 holomorphic Chern-Simons theory will be described as the subsector of physical states of the open superstring which are in the cohomology of the component $q_+$ of
the spacetime supersymmetry generator. Amplitudes will be computed for these states using both the standard RNS prescription and using the topologically twisted prescription. It should be similarly possible to describe D=5 BCOV theory as a subsector of the closed superstring, but this will not be described here.
To allow all fields and pure spinors to be real, it will be convenient to Wick-rotate the $SO(9,1)$ signature to $SO(5,5)$.

Under the $SL(5)$ subgroup of $SO(5,5)$, the 16 components of the D=10 spacetime supersymmetry generator $q_\a$ split into $(q_+, q^{ab}, q_a)$ where $a=1$ to 5,
and the ten components of the D=10 translation generator $P_m$ split into $(P_a, P^a)$. Since $\{q_+, q_a\} = P_a$, any state in the cohomology of $q_+$ is annihilated by $P_a$ and is therefore
massless. The massless onshell states of the open superstring are the D=10 super-Yang-Mills fields $(A_m, \chi^\a)$, and it will now be argued that D=5 holomorphic Chern-Simons states are precisely the D=10 super-Yang-Mills states with $P_a=0$.

The onshell D=10 super-Yang-Mills fields are $(A_m, \chi^\a)$ whose linearized equations of motion and supersymmetry transformations are
\eqn\leom{\p^n F_{mn} =0, \quad \p_m (\g^m \chi)_\a =0, \quad
\d A_m = \e\g_m\chi, \quad \d\chi^\a =\half (\e\g_{mn})^\a F_{mn}}
where $F_{mn} = \p_{[m} A_{n]}.$ So if $P_a=0$, the equations of motion for $(A_a, A^a, \chi^+, \chi_{ab}, \chi^a)$ are
\eqn\eomm{\p^b\p^a A_a = 0, \quad \p^{[a} \chi^{b]} = \p^a \chi_{ab} = 0,}
and the supersymmetry transformation in the direction $q_+$ is
\eqn\susyplus{\d A_a =0, \quad \d A^a = \chi^a, \quad \d \chi^+ = F^a{}_a = \p^a A_a, \quad \d\chi_{ab} = F_{ab} =0, \quad \d \chi^a =0.}

From \eomm\ and \susyplus, one sees that the supersymmetry transformations $\d(\p^a \chi^+)$ and $\d F^{ab}$ are zero. Furthermore, since $\d A^a = \p^a \s$ is a gauge transformation of $A_m$, $\chi^a$ is in the cohomology of $q_+$ when it can be expressed as $\chi^a = \p^a \s$ for some $\s$ .
So the super-Yang-Mills fields in the cohomology of $q_+$ can be conveniently combined into the $SL(5)$-covariant superfield
\eqn\cs{\Phi(x_a, \G^b) = \s(x)  + A_a(x) \G^a + \chi_{ab}(x) \G^a \G^b + F^{ab}(x) (\G^3)_{ab} + \p^a \chi^+(x) (\G)_a^4}
where $\s$ is related to the gluino field by $\chi^a = \p^a \s$ and $\G^a$ are 5 anticommuting variables. In terms of $\G^a$, the other supersymmetry generators are
$q_a = {\p\over{\p\G^a}}$ and $q^{ab} = P^{[a} \G^{b]}$.

\subsec{RNS amplitude prescription}

For these $x^a$-independent states, the super-Yang-Mills three-point amplitude ${\cal A} =Tr \int d^{10} x ( \chi^\a \g^m_{\a\b} \chi^\b A_m + F^{mn} A_m A_n)$ simplifies to
\eqn\treesimp{ {\cal A} =Tr \int d^{10} x ( \{\chi^+, \chi^a \} A_a + \e^{abcde} \chi_{ab}\chi_{cd} A_e + \{\chi_{ab}, \chi^a\} A^b +  F^{ab} A_a A_b)}
where we are not being careful with the relative coefficients.
If one chooses the vertex operators for $(\chi^a, A_a, \chi_{ab}, A^b, \chi_+)$ to carry picture $P= (-{3\over 2}, -1, -\half, 0, \half)$, one can check that all terms in \treesimp\ carry
picture $-2$. Since the RNS tree amplitude prescription requires picture $-2$, there are no picture-changing operators needed for the tree-level computation with this choice of picture.
The corresponding BRST-invariant vertex operator for the 
holomorphic Chern-Simons fields is
\eqn\unint{V_{RNS} =  \s c e^{-{3\over 2}\phi} \S^+  + A_a c e^{-\phi} \psi^a + \chi_{ab} c e^{-\half \phi} \S^{ab} }
$$+ A^a (\g \psi_a + c\p x_a) + F^{ab} c\psi_a\psi_b +
 Q(\xi ce^{ -\half\phi}\S_+ \chi^+)$$
where $\S_\a$ is the usual spin field constructed from $\psi_m$.
Note that the
vertex operator $\s c e^{-{3\over 2}\phi} \S_+$ is related by picture-changing to the usual vertex operator in the $-\half$ picture $\chi^a c  e^{-{1\over 2}\phi} \S_a$ where $\chi^a = \p^a \s$.

The 3-point tree amplitude is easily computed in the RNS formalism using these vertex operators and one finds
\eqn\threepoint{{\cal A} = \langle V_{RNS}^{ (1)} (z_1) V_{RNS}^{ (2)} (z_2) 
V_{RNS}^{ (3)} (z_3)\rangle }
$$ = \d^5(0) \d^5 (\sum k_r^a) Tr[\e^{abcde} \chi^{(1)}_{ab} A^{(2)}_c \chi^{(3)}_{de} + F^{(1)ab} A^{(2)}_a A^{(3)}_b + ...]$$
$$= \d^5(0) \d^5 (\sum k_r^a) \int d^5 \G~Tr [\Phi^{(1)}(\G) \Phi^{(2)} (\G) \Phi^{(3)}(\G)] $$
where $\langle c \p c \p^2 c e^{-2 \phi}\rangle$ is the RNS zero mode measure factor, $\d^5(0)$ comes from the integration over $x^a$ zero modes, and $\Phi(\G)$ is the superfield of \cs.

To compute $N$-point $g$-loop superstring amplitudes of these states, note that with this choice of picture, the kinetic term is
$\langle V_{RNS} Q Y V_{RNS}\rangle$ where $Y= c \p \xi e^{-2\phi}$ is the picture-lowering operator. This implies that each propagator
is accompanied by a picture-raising operator, so one needs to insert $3g-3+N$ picture-raising operators for an $N$-point $g$-loop amplitude whose
amplitude prescription is
\eqn\loop{{\cal A} = \int d^{3g-3+N} \tau \langle \prod_{r=1}^N V_{RNS}^{ (r)} \prod_{n=1}^{3g-3 + N} [ e^{\phi} (\p x^a \psi_a + \p x_a \psi^a) + ... ] (\int \mu_n b)\rangle}
where $\mu_n$ are the $3g-3+N$ Beltrami differentials for a genus $g$ surface with $N$ punctures and $...$ denotes terms in the picture-raising operators independent of $\psi^m$. Choosing $N$ of the picture-raising operators and $b$ ghosts to coincide with the vertex operator locations, one obtains 
\eqn\looptwo{{\cal A} = \int d^{3g-3} \tau \langle \prod_{r=1}^N U_{r} \prod_{n=1}^{3g-3} [ e^{\phi} (\p x^a \psi_a + \p x_a \psi^a) + ... ] (\int \mu_n b)\rangle}
where the integrated vertex operators are 
\eqn\integ{U = \int dz  (\chi^a e^{-\half \phi} \S_a + A_{a} \p x^ a +\p^b A_{a} \psi_b \psi^a + ... ).}

By counting the $GL(1)$ charge of the $\psi^m$ dependence in the vertex operators, it will now be shown that only the term $e^\phi \p x^a \psi_a$
contributes to the picture-changing operators in \loop. Defining $\psi^a$ to carry $+1$ charge and $\psi_a$ to carry $-1$ charge,
the $GL(1)$ charge carried by $\psi^m$ in all terms of the vertex operator $U$ of \integ\
is always greater than or equal to $-3P$ where $P$ is the picture. And since the vertex operators $\prod_{r=1}^N 
U_{r}$ must contribute $P=1-g$
to get a non-vanishing amplitude using the prescription of \loop, the $GL(1)$ charge of the $\psi$'s in the vertex operators must be greater than or equal to
$3g-3$. Therefore, each of the $3g-3$ picture-raising operators must contribute $e^\phi \p x^a \psi_a$ to cancel the $3g-3$ $GL(1)$ charge from the vertex
operators. Note that $e^\phi \p x^a \psi_a$ has no poles with itself or with any of the vertex operators, so the picture-raising operators in \loop\ can be inserted anywhere
on the surface and it will be convenient to insert them at the same locations as the $3g-3$ $b$ ghosts.

To compute \loop, one still has to peform the functional integral over the RNS matter and ghost fields. A convenient trick for avoiding the need to sum over
spin structures is to find a field redefinition which maps the RNS worldsheet variables to Green-Schwarz-like variables with integer spin. One such map is the
$U(5)$ ``hybrid formalism" of \NBufive\NBooguri\ in which fermionic variables $\t^a$ and conjugate momenta variables $p_a$ of conformal weight 0 and 1 are defined as
\eqn\hybrid{ \t^a = e^{\half \phi} \S^a, \quad p_a = e^{-\half \phi} \S_a.}
One can then define $e^\rho = e^{{3\over 2} \phi} \S_+$ and $e^{-\rho} =  e^{-{3\over 2} \phi} \S^+$ which carry conformal weight $-2$ and $+1$ respectively and have no singular OPE's with $\t^a$ and $p_a$.

In terms of these new variables, $e^\phi \p x^a \psi_a = e^\rho \p x^a p_a$ and $U = \int dz  (\chi^a p_a + A_a \p x^ a + \p^b A_a p_b \t^a + ... ).$ If the picture-changing
operators are inserted together with the $b$ ghosts in \loop, the functional integral over $(b,c)$ cancels the functional integral over $\rho$ and \loop\ simplifes to 
\eqn\lsimple{{\cal A} = \int d^{3g-3} \tau \langle \prod_{r=1}^N U_{r} \prod_{n=1}^{3g-3} \int \mu_n \p x^a p_a \rangle.}
This simple expression for D=5 holomorphic Chern-Simons genus $g$ $N$-point amplitudes includes, for example, the well-known D=4 topological open superstring amplitude if one chooses 
$2g-2$ external states as D=4 gluinos $\chi^A$ and two external states as $D=4$ self-dual gluons $A_A$ where we have split the
$a=1$ to 5 index into $A=1$ to 2 and $J=3$ to 5.  With this choice of external states, one finds after integration over the $D=4$ variables $(\t^A, p_A)$ that 
\eqn\ltop{{\cal A} = (\chi^A \chi^B \epsilon_{AB})^{g-1} (F^A_B F_A^B) \int d^{3g-3} \tau \langle\prod_{n=1}^{3g-3} \int\mu_n\p x^J p_J \rangle}
where $F_A^B$ is the self-dual field-strength \NBooguri.

\subsec{ Topological amplitude prescription}

Since the amplitudes computed for these states using the RNS prescription have such a simple form, it should not be surprising that D=5 holomorphic Chern-Simons 
amplitudes can alternatively be computed using a ``topological prescription". As discussed in the introduction, the first step is to perform
a similarity transformation on the vertex operator $V_{RNS}$ of \unint\ satisfying $(Q-q_+^{+\half}) V_{RNS} = \eta_0 V_{RNS}=0$ to a vertex operator $V_{top}= e^{B} e^{A} V_{RNS} e^{-A} e^{-B}$ satisfying $Q V_{top} = q_+^{-\half} V_{top} =0$ where
$A=\oint  e^{-{\phi\over 2}}\S_+ \xi$ and $B = \oint   e^{\phi\over 2}\S^+\eta$. 

Starting with $V_{RNS}$ of \unint, one finds
\eqn\vtop{V_{top} = e^{B} e^{A} V_{RNS} e^{-A} e^{-B}}
$$=e^{B}[ V_{RNS} +c [(\half\xi \p \phi +\half\xi \psi_a\psi^a + \p\xi) e^{- 2\phi} \s + \xi \p \xi e^{-{5\over 2}\phi}\S_+ \s + \xi e^{-{3\over 2}\phi}\S^a A_a]e^{-B}$$
$$=
 \s c \xi \p\xi e^{-{5\over 2}\phi} \S_+  + A_a c \xi e^{-{3\over 2}\phi} \S^a + \chi_{ab} c e^{-\half \phi} \S^{ab} $$
$$+ A^a (\g \psi_a + c\p x_a) + F^{ab} c( \psi_a\psi_b + \eta e^{\half\phi}\S_{ab})$$
$$+ Q [\chi^+  c (\xi e^{-\half \phi} \S_+ + \eta\xi + \half\p\phi + \half \psi^\a \psi_a + \eta e^{\half\phi} \S^+)].$$

The next step is to perform a field redefinition to variables $(\G^a, \Gb_a)$ of conformal weight $(0,1)$
\eqn\defgammatwo{\G^a = \g\psi^a = \eta e^\phi \psi^a, \quad \Gb_a = (\g)^{-1} \psi_a = \xi e^{-\phi} \psi_a,}
and to twist the $(\g,\b)$ ghosts into ``tilded" versions $(\gwt,\bwt)$ of conformal weight $(-1,+2)$ as 
\eqn\tildedghosts{\widetilde \g = (\g)^2 = \eta\p\eta e^{2\phi}, \quad \bwt = (2\g)^{-1}\b + (2\g^2)^{-1}\psi^a \psi_a.} 
Fermionizing $\gwt = \etawt e^{\phiwt}$ and $\bwt = \p\xiwt e^{-\phiwt}$ and requiring that $(\etat, \xit)$ have no poles with
$(\G^a, \Gb_a)$ implies that 
\eqn\fermtilde{\etat= e^{-{\phi \over 2}} \Sigma_+, \quad\xit= e^{{\phi \over 2}} \Sigma^+, \quad e^{\phiwt} = e^{{5\phi}\over 2} \eta\p\eta \Sigma^+, \quad e^{-\phiwt} = e^{-{{5\phi}\over 2}} \xi\p\xi \Sigma_+ .}

In terms of these twisted variables, $V_{top}$ of \vtop\ satisfies $Q V_{top} = \etawt_0 V_{top} =0$ and takes the simple form
\eqn\vtoptw{V_{top} =  c e^{-\phiwt}  \Phi (\G) + Q\Lambda}
where $\Phi(\G) = \s  + A_a \G^a + \chi_{ab} \G^a \G^b + 
\p^a A^b (\G^3)_{ab} + \p^a \chi^+ (\G)_a^4$ is the superfield of \cs, 
\eqn\brstcs{Q =\oint[ \p x_a \G^a + \gwt (b + \p x^a \Gb_a) - c (\p x^a \p x_a + \Gb_a \p \G^a + \bwt\p\gwt + \p (\bwt\gwt) + b\p c )],}
$${\rm and}\quad \Lambda = A^b c \Gb_b + \chi^+ c (\Gb_b \G^b + \xit\etat + 2\p\phiwt +  \etawt e^{2\phiwt} (\Gb)^5).$$  

Since $Q =e^{- \oint (c \p x^a \Gb_a + \bt c \p c)} \oint (\p x_a \G^a + \gt b) e^{ \oint (c \p x^a \Gb_a + \bt c \p c)}$,
states in the cohomology of $Q$ are related to states independent of $(b,c)$ and $(\bt, \gt)$ in the cohomology of $G^+ = \p x_a \G^a$. So as expected for holomorphic Chern-Simons states, $V_{top}$ only depends on the zero modes of $x_a$ and $\G^a$ and is described by the superfield $\Phi(x_a, \G^a)$ of \cs. 

The initial RNS amplitude prescription is 
\eqn\rnsin{{\cal A} = \int d^{3g-3+N} \tau \langle \prod_{r=1}^N V_{RNS}^{(r)} \prod_{n=1}^{3g-3 + N} Q(\xi) (\int \mu_n b)\rangle}
where $V_{RNS}$ is defined in \unint, 
and \rnsin\ can be expressed as
\eqn\rnstwo{{\cal A} = \int d^{3g-3+N} \tau \langle \prod_{r=1}^N V_{RNS}^{(r)} \prod_{n=1}^{3g-3 + N} (Q(\xi)-q_+^{ +\half}(\xi)) (\int \mu_n b)\rangle}
since the terms $q_+^{+\half}(\xi)$ in \rnstwo\ do not contribute because of ghost-number conservation.
After the similarity transformation of \vtop, \rnstwo\ becomes
\eqn\becomes{
{\cal A} = \int d^{3g-3+N} \tau \langle \prod_{r=1}^N (c e^{-\phiwt} \Phi^{(r)} (x,\G) + Q\Lambda_r) \prod_{n=1}^{3g-3 + N} Q(\xi+\xit) (\int \mu_n b)\rangle}
where $e^B e^A \xi e^{-A} e^{-B} = \xi+\xit$, $e^B e^A \eta e^{-A} e^{-B} = \etat$  and $e^B e^A (Q-q_+^{+\half}) e^{-A} e^{-B} = Q$ have been used. In terms of the tilded variables, $\xi = e^{-2\phit- H}$ which anticommutes with $\etat_0$, so $Q(\xi)$ is BRST-trivial in the small tilded Hilbert space. This means that the terms $Q(\Lambda_r)$ and $Q(\xi)$ only contribute surface terms in \becomes\ which will be ignored. After choosing the $3g-3+N$ picture-changing operators to be located on top
of the $b$ ghosts, one obtains
\eqn\becomesb{
{\cal A} = \int d^{3g-3+N} \tau \langle \prod_{r=1}^N (c e^{-\phiwt} \Phi^{(r)} (x,\G) ) \prod_{n=1}^{3g-3 + N}  (\int \mu_n b e^{-\phit} \p x^\a \Gb_a).\rangle}
Finally, after integrating out the $(b,c)$ ghosts and $(\bt, \gt)$ whose correlation functions cancel, one obtains the
$N$-point $g$-loop topological amplitude prescription 
\eqn\looptop{{\cal A} = \int d^{3g-3} \tau \langle \prod_{r=1}^N \Phi^{(r)}(x, \G) \prod_{n=1}^{3g-3+N} \int \mu_n \p x^a \Gb_a\rangle .}

\newsec{B-RNS-GSS formalism}

In this section, the N=1 worldsheet supersymmetric action of the B-RNS-GSS formalism will be described for the open superstring where the right-moving and auxiliary variables in the worldsheet superfields will be ignored. In addition to the spacetime 
vector worldsheet superfield $X^m = x^m + \k \psi^m$ of the RNS formalism where $\k$ is an anticommuting parameter, this formalism includes the spacetime spinor worldsheet superfields $\T^\a = \t^\a + \k \L^\a $ and 
$\Phi_\a = \O_\a + \k p_\a$, where $\t^\a$ and its conjugate momentum $p_\a$ are fermionic Green-Schwarz-Siegel variables of conformal weight $(0,1)$, and $\L^\a$ and its conjugate
momentum $\O_\a$ are bosonic twistor-like variables of conformal weight $(\half, \half)$. 

To cancel the conformal anomaly contribution of $(\T^\a, \Phi_\a)$, the B-RNS-GSS formalism also includes the spacetime spinor bosonic worldsheet superfield $\Vb_\a= -\Lb_\a + \k R_\a$ and its fermionic conjugate momentum superfield
$\Wb^\a = S^\a + \k \Ob^\a$ where $(\Lb_\a, \Ob^\a)$ carries conformal weight $(0,1)$ and $(R_\a, S^\a)$ carries conformal weight $(\half,\half)$. These superfields
will later be related to the non-minimal superfields in the pure spinor formalism.



In terms of these superfields, the worldsheet action is 
\eqn\actiont{S = -\int d^2 z \int d^2 \k [ \half D X^m \overline D X_m + \Phi_\a \overline D \T^\a + 
\Wb^\a \overline D\Vb_\a ]  }
where $D = {\p\over{\p\k}} - \k \p_z$ and $\overline D ={\p\over{\p\overline\k}} + \overline\k \p_{\overline z}$ are the fermionic N=(1,1) derivatives.
The action of \actiont\ is manifestly invariant under worldsheet superconformal invariance and 
the N=1 superconformal generator is 
\eqn\scfg{{\cal G} =DX^m \p X_m - D \Phi_\a D\T^\a - \Phi_\a \p\T^\a - D\Wb^\a D\Vb_\a + \Wb^\a \p\Vb_\a  .}
In addition to manifest worldsheet supersymmetry, this action and N=1 superconformal generator are also invariant under the $D=10$ spacetime supersymmetry transformations 
\eqn\susyonet{\d\T^\a = \e^\a, \quad \d X^m = -\half \e\g^m \T, \quad \d \Phi_\a = \half (\e\g^m)_\a DX_m - {1\over 8} (\e\g^m \T)(\g_m D\T)_\a,}
generated by
\eqn\susyq{q_\a = \int dz\int d\k [ \Phi_\a -\half  D X^m (\g_m\T)_\a +{1\over{24}} (\T\g^m D\T) (\g_m \T)_\a]}
and which satisfies the algebra $\{q_\a, q_\b\} = -\g^m_{\a\b}\int dz \int d\k  ~DX_m = \g^m_{\a\b}\int dz ~\p x_m$. 

After setting auxiliary fields in the worldsheet superfields to zero and ignoring the right-moving variables, the action of \action\ in terms of components is
\eqn\actioncomp{S = \int d^2 z [\half \p x^m \pb x_m +\half \psi^m \pb \psi_m +  p_\a \pb \t^\a + \O_\a \pb \L^\a + \Ob^\a\pb \Lb_\a + S^\a \pb R_\a 
] }
and the
spacetime supersymmetry generator of \susyq\ is 
\eqn\susycomp{ q_\a = \oint [ p_\a +  \half\p x^m (\g_m\t)_\a + {1\over 24} (\t\g^m \p\t) (\g_m \t)_\a + \half\psi^m (\g_m \L)_\a +{1\over{16}} (\L\g^m\L)(\g_m \t)_\a].}
In components, the N=1 superconformal generator ${\cal G}= G + 2\k T$ of \scfg\ is
\eqn\comg{G = \psi^m \p x_m - \L^\a p_\a -  \O_\a \p\t^\a - \Ob^\a R_\a - S^\a \p \Lb_\a
, }
\eqn\stresst{ T = -\half \p x^m \p x_m - p_\a \p\t^\a  - \half \psi^m \p \psi_m - \half (\O_\a \p \L^\a - \L^\a \p \O_\a ) }
$$ - \half (R_\a \p S^\a + S^\a \p R_\a) - \Ob_\a \p\Lb^\a.$$
 

To make spacetime supersymmetry manifest in this formalism, one performs the similarity transformation 
\eqn\simsusy{{\cal O} \to e^{\half\oint \psi^m (\L \g_m\t)}
{\cal O}  e^{-\half\oint \psi^m (\L \g_m\t)}}
 on all operators ${\cal O}$, which transforms the spacetime supersymmetry generator of \susycomp\ to the Green-Schwarz-Siegel supersymmetry generator of the pure spinor formalism
\eqn\susygss{ q_\a = \oint [ p_\a + \half \p x^m (\g_m\t)_\a + {1\over 24}(\t\g^m \p\t) (\g_m \t)_\a]. } 
Although the similarity transformation of \simsusy\ breaks the manifest worldsheet supersymmetry since $[G,\oint \psi^m (\L\g_m\t)]$ is non-zero, it preserves the action and stress tensor $T$ of \actioncomp\ and \stresst\ and transforms 
the N=1 superconformal generator of \comg\ into
\eqn\stresstsusy{ G = \psi^m \pi_m - \L^\a d_\a + \half(\L\g^m \L) \psi_m -  \O_\a \p\t^\a - \Ob^\a R_\a - S^\a \p \Lb_\a }
where $d_\a = p_\a -\half \p x_m (\g^m\t)_\a -{1\over 8} (\t\g_m \p\t) (\g^m\t)^\a$
and $\pi^m = \p x^m + \half\t\g^m \p\t$ are the spacetime supersymmetric variables of the Green-Schwarz-Siegel formalism. Note that $T$ of \actioncomp\ can be expressed in manifestly spacetime supersymmetric form as 
\eqn\stresss{ T = -\half \pi^m \pi_m - d_\a \p\t^\a  - \half \psi^m \p \psi_m - \half (\O_\a \p \L^\a - \L^\a \p \O_\a ) }
$$ - \half (R_\a \p S^\a + S^\a \p R_\a) - \Ob_\a \p\Lb^\a.$$
 
Putting together the critical N=1 superconformal generators of \stresstsusy\ and \stresss\ with the
$(b,c; \beta, \gamma)$ N=1 superconformal ghosts, the BRST operator in the B-RNS-GSS formalism is defined as
\eqn\brstb{Q = \oint [  \g G  + c (T -{3\over 2} \b\p\g -\half \p \b\g - b \p c) + \g^2 b]}
where the central charge of $T$ is cancelled as usual by the contribution of the $N=1$ superconformal ghosts so that $Q^2 =0$. 

Because of the extra fields in the B-RNS-GSS formalism, vertex operators $V$ in the cohomology of $Q$ include extra states in addition to the usual physical RNS superstring states. As in holomorphic Chern-Simons, these extra states can be removed by requiring physical states to be annihilated
by $\etat = e^{- \half (\phi + \int^z J)}$ where $J$ is a $U(1)$ generator which commutes with $G$ to form an $N=2$ superconformal field theory. As shown in the following subsection, the
resulting $N=2$ superconformal field theory after twisting is the non-minimal pure spinor formalism.

\subsec{ Relation with pure spinor formalism}

To relate the $N=1$ worldsheet supersymmetric B-RNS-GSS formalism with the twisted N=2 worldsheet supersymmetric
pure spinor formalism, one needs to construct a $U(1)$ current $J$ satisfying $[J, G] = G^+ - G^-$ where $[T, G^+, G^-, J]$
satisfy an $N=2$ superconformal algebra. The first step to constructing $J$ is to express the unconstrained bosonic spinors
$(\L^\a, \Lb_\a)$ of the B-RNS-GSS formalism in terms of the constrained spinors $(\l^\a, \lb_\a)$ of the non-minimal pure spinor formalism
where $(\l^\a, \lb_\a)$ satisfy the pure spinor constaints
\eqn\purel{\l\g^m \l =\lb\g_m\lb=0.}
The relation between $(\L^\a, \Lb_\a)$ and $(\l^\a, \lb_\a)$ is given by
\eqn\rellambda{\L^\a = \l^\a + u^m {{(\g_m \lb)^\a}\over{2(\l\lb)}}, \quad 
\Lb_\a = \lb_\a + \ub_m {{(\g^m \l)_\a}\over{2(\l\lb)}} }
where $(u^m, \bar u_m)$ are spacetime vectors which each have 5 independent components.
It will also be useful to express the unconstrained fermionic spinor $R_\a$ of the B-RNS-GSS formalism in terms of the constrained fermionic spinor $r_\a$ satisfying $(r\g^m \lb)=0$ of the non-minimal pure spinor formalism through the relation
\eqn\relar{R_\a = r_ \a + \rho_m {{(\g^m \l)_\a}\over{2(\l\lb)}} }
where $\rho_m$ has 5 independent components. 

Defining $(w_\a, v_m, \wb^\a, \vb^m)$ to be the bosonic conjugate momenta of $(\l^\a, u^m, \lb_\a, \ub_m)$ and 
$(s^\a, \tau^m)$ to be the fermionic conjugate momenta of $(r_\a, \rho_m)$, the constraints on the non-minimal pure spinor variables are 
\eqn\constotal{\l\g^m\l =\lb\g^m \lb = r \g^m \lb =0, \quad v_m (\g^m \lb)^\a = \ub_m (\g^m\lb)^\a = \rho_m (\g^m\lb)^\a + \ub_m (\g^m r)^\a =0,}
which imply gauge invariances of the conjugate variables. These constraints and gauge invariances imply that
the variables $(\l^\a, u^m, \lb_\a, \ub_m,  r_\a, \rho_m;$
$w_\a, v_m, \wb^\a, \vb^m, s^\a, \tau^m)$ contain the same number of independent components as $(\L^\a, \Lb_\a, R_\a; \O_\a, \Ob^\a, S^\a)$.

The next step to constructing $J$ is to require that all variables of conformal weight $\half$ carry $\pm 1$ charge with respect to $J$. Choosing $(\L^\a, R_\a, {{(\l\g^m\g^n\lb)}\over{2(\l\lb)}}\psi_n)$ to carry charge $+1$ and $(\O_\a, S^\a, {{(\lb\g^m\g^n\l)}\over{2(\l\lb)}}\psi_n) $ to carry charge $-1$ would imply that $J = -\L^\a \O_\a + R_\a S^\a - {{(\l\g^m \g^n \lb)}\over {2(\l\lb)}} \psi^m \psi^n$. However, 
this $U(1)$ generator does not satisfy $[J, G] = G^+ - G^-$ because of the dependence on non-minimal variables and one needs to add an additional term $-{{(\lb\g^m \g^n\g^p r)}\over {4(\l\lb)^2}} \psi_m \psi_n\psi_p$ to $J$ in order to form the desired $N=2$ algebra. So the final expression for the $U(1)$ generator of the $N=2$ algebra is
\eqn\brnsj{ J = -\L^\a \O_\a + R_\a S^\a - {{(\l\g^m \g^n \lb)}\over {2(\l\lb)}} \psi^m \psi^n - {{(\lb\g^m \g^n\g^p r)}\over {4(\l\lb)^2}} \psi_m \psi_n\psi_p}
$$= -\L^\a \O'_\a + R_\a S'^\a + \G^m \Gb_m$$ 
where $(\G^m, \Gb_m)$ is defined in terms of $\psi^m$ as 
\eqn\defgamma{ \G^m = {{(\l\g^m \g^n \lb)}\over {2(\l\lb)}} \psi_n, \quad
\Gb_m = {{(\lb\g_m \g_n \l)}\over {2(\l\lb)}} \psi^n + {{(\lb\g_m \g_n\g_p r)}\over {8(\l\lb)^2}}\psi^n\psi^p}
and $(\O'_\a, \Ob'^\a, S'^\a)$ are defined to have no poles with $(\G^m, \Gb_m)$, i.e. 
\eqn\nopol{\int d^2 z (\O'_\a \pb \L^\a + \Ob'^\a \pb \Lb_\a + S'^\a \pb R_\a + \Gb_m \pb\G^m ) = }
$$
\int d^2 z (\O_\a \pb \L^\a + \Ob^\a \pb \Lb_\a + S^\a \pb R_\a + \half \psi_m \pb\psi^m) .$$
Note that the $(\G^m, \Gb_m)$ variables carry charge $(+1, -1)$ and are constrained to satisfy $\Gb_m (\g_m \lb)^\a =0$, which together with the gauge invariances implied by this constraint, imply that $(\G^m, \Gb_m)$ contain the same number of independent components as $\psi^m$.

In terms of these variables, the $N=2$ fermionic superconformal generators are
\eqn\gplus{G^+ = -\L^\a d_\a -\Ob^\a R_\a + \half (\L\g^m\L)\psi_m + {{(\l\g^m \g^n \lb)}\over {2(\l\lb)}}\pi_m \psi_n +  Y,}
$$= e^U (-\l^\a d_\a +u^m \Gb_m -  {\Ob'}^\a R_\a ) e^{-U},$$
$$G^- = -\O_\a \p\t^\a - S^\a \p\Lb_\a + {{(\l\g^m \g^n \lb)}\over {2(\l\lb)}}\psi_m \pi_n-Y,$$
$$= -\O'_\a \p\t^\a - S'^\a \p\Lb_\a + \pi^m \Gb_m -{{(\l\g_m\g_n r)}\over{4(\l\lb)}}\Gb_m\Gb_n$$
 where 
 \eqn\simRR{U = \oint \G^m [{ {(\lb \g_m d)}\over{2(\l\lb)}} 
 +{{(\lb\g^{mnp}r)}\over{16(\l\lb)^2}} 
  (\O'\g_{np}\L)] \quad{\rm and}}
$$Y = [ {{(\l\g^m \g^n r)}\over {4(\l\lb)}}-{{(\l\g^m \g^n \lb)(\l r)}\over {4(\l\lb)^2}}+{{(\p\t\g^m \g^n \lb)}\over {4(\l\lb)}}-
{{(\l\g^m \g^n \lb)(\lb\p\t)}\over {4(\l\lb)^2}}  
-{{(\lb\g^m \g^n\g^p r)\pi_p}\over {8(\l\lb)^2}} ]\psi_m \psi_n$$
$$+[{{(\lb\g^m \g^n \g^p \p\lb)}\over {24(\l\lb)^2}}- {{(\lb\g^m \g^n \g^p r)(\lb\p\t)}\over {12(\l\lb)^3}}+ {{(r\g^m \g^n \g^p r)}\over {24(\l\lb)^2}}- {{(\lb\g^m \g^n \g^p r)(\l r)}\over {12(\l\lb)^2}}]\psi_m\psi_n\psi_p.$$

The next step to relate the B-RNS-GSS formalism to the pure spinor formalism is to twist the $N=2$ superconformal generators
by shifting $T \to T + \half \p J$ and replacing the $(\b, \g)$ ghosts of conformal weight $({3\over 2}, -\half)$ with the
$(\bt, \gt)$ ghosts of conformal weight $(2, -1)$. As in the $D=5$ holomorphic Chern-Simons theory of the previous section,
 $\gt = (\g)^2$ and $\bt = (2\g)^{-1} \b +(2\g^2)^{-1} J$ implies that $\gt = \etat e^{\phit}$ and $\bt = \p \xit e^{-\phit}$
where
\eqn\etadefin{ \etat = e^{- \half (\phi + \int^z J)} = e^{-{\phi\over 2}} (:\l^\a \S_\a: -(\l\g_{mnp} \S){ {\lb\g^{mnp} r}
\over {8(\l\lb)^2}}),}
$$\xit =  e^{\half(\phi +\int^z J)} = e^{{\phi\over 2}} :w_\a \Sb^\a:.$$
The spin fields $\S_\a$ and $\Sb_\a$ of \etadefin\ are defined to contain square-root cuts with $(\L^\a, R_\a, \O_\a, S^\a)$ in addition to the usual square-root cuts with $\psi^m$, i.e. 
\eqn\defsigmaa{\S_\a = \S_\a^{RNS} ~e^{\half \oint^z (-\L^\b \O_\b + R_\b S^\b)}, \quad
\Sb_\a = \S_\a^{RNS}~ e^{-\half \oint^z (-\L^\b \O_\b + R_\b S^\b)}
.} 
In \etadefin, $:\l^\a \S_\a:$ denotes the residue of the square-root pole in the OPE of $\l^\a$ with $\S_\a$ and $:w_\a \Sb^\a:$ denotes the residue of
the square-root pole in the OPE of $w_\a$ with $\Sb^\a$.
Note that both $\S_\a$ and $\Sb_\a$ carry $5\over 8$ conformal weight because $(-\L^\b \O_\b + R_\b S^\b)$ has no singularity with itself.

This twisting corresponds to redefining
\eqn\twistcs{ \L^\a \to \g^{-1} \L^\a,  \quad R_\a \to \g^{-1} R_\a, \quad \G^m \to \g^{-1} \G^m, }
$$\O'_\a \to \g \O'_\a, \quad S'^\a \to \g S'^\a, \quad \Gb_m \to \g \Gb_m$$
so that $(\L^\a, R_\a, \G^m)$ carry zero conformal weight and $(\O'_\a, S'^\a, \Gb_m)$ carry $+1$ conformal weight.
In terms of the twisted variables, $G^+$ and $G^-$ of equations \gplus\  now carry conformal weight 1 and 2 respectively, and the BRST operator of \brstb\ is mapped to
\eqn\newbrst{ Q = \oint  [ G^+ + \gt (b+G^-) + c (T+\half\p J) + c(\p c b - 2\bt\p\gt-\gt\p\bt)].}

Finally, performing the similarity transformation
\eqn\simbrst{Q \to e^{\oint (c G^- + \bt c \p c) } e^{-U} Q  e^{U} e^{ -\oint (c G^- + \bt c \p c)}}
where $U$ is defined in \simRR, the B-RNS-GSS BRST operator is mapped to the non-minimal pure spinor BRST operator
\eqn\nonminbrst{Q = \oint  (-\l^\a d_\a + u^m \Gb_m -\wb^\a r_\a - \vb^m \rho_m + b\gt)}
where the term  $\oint  (u^m \Gb_m -\wb^\a r_\a - \vb^m \rho_m +b\gt)$ implies that the BRST cohomology is independent of the
non-minimal variables $(u^m, v_m, \G^m, \Gb_m, \wb^\a, \lb_\a, \ub^m, \vb_m, \rho_m, \tau^m)$ and the ghosts
$(b,c,\bt,\gt)$.
Furthermore, in the gauge $\Gb=0$,  $e^{-U} G^-  e^{U}$ is equal to the  
pure spinor composite $B$ ghost \NBdynamical
\eqn\compb{B = 
 - S'^\a\p\Lb_\a  + {{\lb_\a (2
\pi^m (\g_m d)^\a-  N_{mn}(\g^{mn}\p\t)^\a
- J \p\t^\a)}\over{4(\lb\l)}} }
$$- {{(\lb\g^{mnp} r)(d\g_{mnp} d +24 N_{mn}\pi_p)}\over{192(\lb\l)^2}}
+ {{(r\g_{mnp} r)(\lb\g^m d)N^{np}}\over{16(\lb\l)^3}} -
{{(r\g_{mnp} r)(\lb\g^{pqr} r) N^{ mn} N_{qr}}\over{128(\lb\l)^4}}$$
with $N_{mn} = -\half w\g_{mn}\l$ and $J = -w\l$.

So using the same arguments as in the previous section for holomorphic Chern-Simons theory, vertex operators in the B-RNS-GSS formalism which are in the cohomology of $Q$ and annihilated by $\etat_0$ of \etadefin\ are related by a similarity transformation
to vertex operators in the non-minimal pure spinor formalism. In principle, these arguments also imply the equivalence of RNS multiloop amplitudes computed using the $N=1$ worldsheet supersymmetric prescription with pure spinor multiloop amplitudes computed
using the topological $N=2$ worldsheet supersymmetric prescription. However, there are at least four issues which need to be better
understood before claiming an equivalence proof of RNS and pure spinor multiloop amplitude prescriptions.

The first issue involves the explicit construction of physical vertex operators in the B-RNS-GSS formalism which are related by a similarity transformation to physical vertex operators in the pure spinor formalism. In \NBcov, physical B-RNS-GSS vertex operators were constructed for super-Yang-Mills which have non-positive charge with respect to the $U(1)$ generator $-\L^\alpha \O_\alpha$. But states corresponding to 
physical vertex operators in the pure spinor formalism need to have non-positive charge with respect to the $U(1)$ generator of \brnsj,
$J =  -\L^\a \O'_\a + R_\a S'^\a + \G^m \Gb_m$, so that $\etat = e^{- \half (\phi + \int^z J)}$ has no poles with the vertex operator. This implies that one needs to add BRST-trivial terms depending on non-minimal variables to the vertex operator of \NBcov, and it would be interesting to explicitly construct these non-minimal terms.

A second issue is that the B-RNS-GSS $N=2$ superconformal field theory has central charge $\hat c =5$ whereas the pure spinor $N=2$ superconformal field theory of \topological\ has
central charge $\hat c=3$. The difference comes from the relation of the B-RNS-GSS 
unconstrained spinor variables $(\L^\a, \O'_\a)$ with the 
constrained pure spinor and vector variables
$(\l^\a, w_\a, u^m, v_m)$. Although both involve 32 independent bosons, the OPE of the current $(-\L^a\O'_\a)$ with the stress-energy tensor $T$ has a triple pole with coefficient $-16$, whereas the triple pole in the OPE of $(-\l^\a w_\a)$ with $T$ has coefficient $-8$ and of 
$(-2u^m v_m)$ with $T$ has coefficient $2 \times (-5) = -10$. Furthermore, the triple pole in the OPE of $(R_\a S'^\a + \G^m \Gb_m)$ with $T$ has coefficient $+16+5=21$ So the total coefficient in the triple pole in the OPE of $J$ with $T$ is $-16+21=+5$ for the B-RNS-GSS formalism and $-8-10+21=+3$ for the pure spinor formalism, which implies the different values of $\hat c$. In order to prove equivalence of the scattering amplitudes in the two formalisms,
one needs to better understand how this difference in central charge affects the prescriptions.




The third issue is that since $\etat\xi$ and $\xit\eta$ carry non-zero picture, they can
have spurious poles on surfaces with non-zero genus. So
the similarity transformations $A=\oint \etat\xi$ and $B= \oint  \xit\eta$ of \simAA\ may get contributions from these spurious poles, 
which would spoil the argument that $\langle e^{B} e^A (...) e^{-A} e^{-B} \rangle = \langle ... \rangle$ for multiloop amplitudes.
Perhaps one can use methods similar to the ``vertical integration" method of \vertical\ to avoid contributions from these spurious poles, but this needs to
be investigated.


Finally, the fourth issue is that although $Q(\xi)$ is BRST-trivial in the small tilded Hilbert space, it contains inverse powers of $(\l\lb)$. As discussed in \topological, BRST-trivial
quantities with inverse powers of $(\l\lb)$ may contribute to scattering amplitudes
if the degree of the inverse power in the correlation function is more divergent than
$(\l\lb)^{-10}$. For the special case of spacetime supersymmetric $F$-terms, it
was shown in \berknek\ that the degree of inverse power in $(\l\lb)$ is always equal or less
than $(\l \lb)^{-10}$. So for these terms, it is expected that $Q(\xi)$ will not
contribute and the B-RNS-GSS amplitude prescription reduces to the pure spinor
amplitude prescription. But for generic $D$-term contributions to multiloop amplitudes,
the inverse powers of $(\l\lb)$ in the integrand may be greater than $(\l\lb)^{-10}$
and the terms $Q(\xi)$ in the B-RNS-GSS prescription may contribute. An important question is
if these contributions coming from $Q(\xi)$ in the B-RNS-GSS prescription can be related to the complicated
regulator of \berknek\ necessary for $D$-term contributions in the pure spinor multiloop amplitude prescription.

\vskip 10pt
{\bf Acknowledgements:}
I would like to thank Nikita Nekrasov for useful discussions and 
CNPq grant 311434/2020-7
and FAPESP grants 2016/01343-7, 2021/14335-0, 2019/21281-4 and 2019/24277-8 for partial financial support.

\listrefs

\end